\documentclass[12pt]{article}
\usepackage{epsfig}
\usepackage{amssymb}
\usepackage{amsmath}

\newcommand{\om}{\omega}
\newcommand{\al}{\alpha}
\newcommand{\ep}{\epsilon}

\newcommand{\lb}{\lbrack}
\newcommand{\rb}{\rbrack}

\newcommand{\msc}[1]{\mbox{\scriptsize #1}}
\newcommand{\dsp}{\displaystyle}

\newcommand{\br}{\mathbb{R}}

\newcommand{\bsz}{\msc{{\bf Z}}}

\newcommand{\cO}{{\cal O}}
\newcommand{\cN}{{\cal N}}
\newcommand{\cM}{{\cal M}}

\renewcommand{\th}{{\theta}}

\newcommand{\be}{\begin{eqnarray}}
\newcommand{\ee}{\end{eqnarray}}
\newcommand{\nn}{\nonumber\\}
\newcommand {\eqn}[1]{(\ref{#1})}
\newcommand {\rmd} {{\rm d}}

\makeatletter
\@addtoreset{equation}{section}

\makeatother

\begin{document}
\rightline{SNUST 060402} \rightline{UT-06-06} \rightline{\tt
hep-th/0605013}
\vskip1cm
%
\centerline{\bf \Large Unitarity Meets Channel-Duality}
 \centerline{\bf \Large for}
 \centerline{\bf \Large Rolling/Decaying D-Branes}
\vskip1cm
\centerline{\bf Yu Nakayama${}^{a}$\footnote{\tt
nakayama@hep-th.phys.s.u-tokyo.ac.jp}, Soo-Jong
Rey${}^{b}$\footnote{\tt sjrey@snu.ac.kr}, Yuji
Sugawara${}^{a}$\footnote{\tt sugawara@hep-th.phys.s.u-tokyo.ac.jp}}
\vskip0.75cm
\centerline{\sl ${}^{a}$ Department of Physics, University of Tokyo}
\centerline{\sl 7-3-1 Hongo, Bunkyo-ku, Tokyo 113-0033 {\rm JAPAN}}
\vskip0.3cm \centerline{\sl ${}^{b}$ School of Physics and Astronomy
\& BK-21 Physics Division} \centerline{\sl Seoul National
University, Seoul 151-747 {\rm KOREA}}
\vskip0.5cm
\vskip1.2cm \centerline{\bf abstract} \vskip0.5cm
Investigations for decay of unstable D-brane and rolling of
accelerated D-brane dynamics have revealed that various proposed
prescriptions give different result for spectral amplitudes and
observables. Here, we study them with particular attention to
unitarity and open-closed channel duality. From {\sl ab initio}
derivation in the open string channel, both in Euclidean and
Lorentzian worldsheet approaches, we find heretofore overlooked
contribution to the spectral amplitudes and obervables. The
contribution is fortuitously absent for decay of unstable D-brane,
but is present for rolling of accelerated D-brane. We finally show
that the contribution is imperative for ensuring unitarity and
optical theorem at each order in string loop expansion.
%
\newpage
\rightline{\sl The shortest path between two truths in the real
domain} \rightline{\sl passes through the complex domain. --- {\rm
J. Hadamard}}
\section{Introduction}
In Lorentz invariant quantum field theory, two of the most
fundamental properties are locality and unitarity. The locality is
necessary for the theory to obey causality. In turn, the
microcausality, expressed in the commutativity of local operators at
spacelike separations, leads to {\sl analyticity} of scattering
amplitudes and analytic continuation thereof. The unitarity is
necessary for the theory to admit meaningful probabilistic
interpretation. In turn, requiring unitarity to the scattering
amplitudes, one obtains information concerning discontinuities about
their branch points~\footnote{In fact, through dispersion relations,
it puts a powerful constraint on the analytic functions.}. In
particular, Feynman's $+i \varepsilon$ prescription on propagators
together with Cutkovsky-Landau cutting rules provides a simple
account for the causality and the unitarity. The Wick rotations and
optical theorem are the simplest consequences of such \cite{QFT}.

In perturbative string theory, it was established that string
scattering amplitudes are Lorentz invariant, unitary and local
despite extended nature of string. Unlike quantum field theory,
however, unitarity and locality are not manifest in the rules of
perturbation theory. Instead, the unitarity was established only
indirectly by showing equivalence between the covariant and the
light-cone formulations of scattering amplitudes. Also, the locality
was proven only by resorting to string field
theory~\cite{Erler:2004hv}. To date, establishing analyticity and
analytic continuations of multi-loop scattering amplitudes have
largely remained an outstanding unsolved problem in string theory.
The issue is somewhat more involved than quantum field theory by the
built-in channel duality between $s$- and $t$-channels, which
exchanges the open and the closed string channels for string loop
amplitudes.

In recent years, numerous works addressed string dynamics in a
variety of time-dependent background. It includes timelike Liouville
theory \cite{TL}, strings in null orbifolds \cite{NO} or
cosmological backgrounds \cite{cosmo}, open string dynamics in
electric field \cite{schwinger} or various time-dependent backgrounds \cite{otd},
decay of unstable D-branes \cite{tachyon} and rolling of accelerated D-branes
\cite{rolling} in the NS5-brane backgrounds. Each of these
investigations involved one way or another certain prescription of
analyticity and analytic continuation of scattering amplitudes
involving string states and D-branes. Those prescriptions were
largely case-specific and often exploited analytic continuations of
not only spacetime variables but also parameters defining the string
worldsheet dynamics. Thus, it was far from transparent whether such
prescriptions are mutually consistent and, after all,
correct~\footnote{Another radiative process where the channel
duality and the optical theorem were investigated is the absorption
of $p$ F-strings by $q$ D-string into $(p,q)$ string bound-state
\cite{leerey}. For rolling dynamics of D-brane accelerated by
F-strings, see \cite{bakreyyee}.}.

The purpose of this work is to bring out certain consistency checks
of the prescription we proposed recently in the context of rolling
D-branes in NS5-brane backgrounds \cite{NST,NPRT,NRS} (see also
\cite{Sahakyan,CLS}), and compare critically with different
prescriptions put forward by other works in this context \cite{OR}
and the decay of unstable D-branes \cite{LLM,KLMS,KMS}. The central
issue is whether analytic continuation can be prescribed in
transition amplitudes of these processes in a way the optical
theorem is manifest and right. Indeed, we shall show that certain
prescriptions that arose from one context does not lead to
self-consistent results when applied to other contexts. We shall
also show that some other prescriptions adopted in the literatures
are inconsistent and incorrect.

As said, we shall address the questions primarily in the context of
rolling of accelerated D-brane in NS5-brane background
\cite{rolling} and of decay of unstable D-brane (either in flat
space \cite{LLM}, in linear dilaton background \cite{KLMS} or in
two-dimensional string theory \cite{KMS}). Both situations involve
conversion of the energy stored in the D-brane to elementary string
states. In the process, formally, the optical theorem facilitates to
extract emission spectra of the decaying or rolling D-brane from
forward scattering amplitude of the D-brane. In string perturbation
theory, the leading order contribution comes from the cylinder
amplitude. Typically, the amplitude is ill-defined and requires
careful prescription.

In the previous works \cite{NST,NPRT,NRS}, we studied rolling
dynamics of accelerated D-brane in NS5-brane background and
two-dimensional black hole geometry, described by ${\cal N}=2$
superconformal SL$(2,\mathbb{R})/$U(1) model. We extracted cylinder
amplitude in Lorentzian worldsheet and Lorentzian spacetime,
respectively, via $+i \varepsilon$ prescriptions. The prescriptions
were devised to render the modular integral (Schwinger-Feynman
integral) of the amplitude well-defined and analytic. In particular,
after integrating over the worldsheet modulus, the amplitude
expressed in the closed string channel develops an imaginary part
corresponding to total emission of on-shell closed string states,
thus ensuring the optical theorem to hold. The total radiation
number was given in the closed string channel by \cite{NST,NPRT,NRS}
\be \overline{\cal N} = \sum_M \sqrt{\rho^{(c)}(M)} \int_0^\infty
{\rmd p \over 2\omega} {\sinh {2 \pi  \over
Q}\sqrt{\frac{\alpha'}{2}}p \over \left[ \cosh {2 \pi \over Q}
\sqrt{\frac{\alpha'}{2}} \omega+ \cosh {2 \pi \over Q
}\sqrt{\frac{\alpha'}{2}} p \right] \sinh \pi Q
\sqrt{\frac{\alpha'}{2}} p}~, \nonumber \ee
where $\rho^{(c)}(M)$ is density of states and $\omega \equiv \sqrt
{p^2 + M^2}$ is on-shell energy of the left-right symmetric closed
string states.

Utilizing the open-closed channel duality, spectral amplitudes and
observables ought to be re-expressible in the open string channel
via (generalized) Fourier transform. The procedure is not always
straightforward and, as we will see in this work, requires careful
treatment of various Fourier transformations involved. In this work,
we undertake {\sl ab initio} analysis of the spectral amplitudes and
observables and find that the Fourier transformations are
generically not convergent. They require suitable analytic
continuations and we propose a specific prescription for how to do
so. With the prescription, we find that both spectral amplitudes and
observables contain new contribution from the analytic continuation
in addition to naive contribution. So, for the total emission number
$\overline{\cal N}$ and the cylinder amplitude $Z_{\rm cylinder}$,
the results take schematically the form:
\be \overline{\cal N} &=& \int_0^\infty {\rmd t_o \over t_o} \Big(
F_{\rm naive}(t_o) + F_{\rm pole}(t_o) \Big) \nn
Z_{\rm cylinder} &=& Z_{\rm naive} + Z_{\rm pole}~, \ee
viz. sum of the naive part and the new contribution part. The naive
part is the result obtained based on a tacit assumption that the
Fourier transforms between open and closed string channels are
always convergent.

Moveover, the analytic continuations we propose are not only
mathematically correct but also physically justified: the naive part
do not obey the optical theorem, whereas sum of the naive and the
new parts do so. Stated differently, consistency between the
open-closed channel duality and the unitarity in string perturbation
theory require the Fourier transform to be defined via the analytic
continuations proposed in this work. We believe the results in this
work clarify much of confusion in previous works concerning spectral
amplitudes and observables for the decay of unstable D-brane and the
rolling of accelerated D-brane. In particular, it shows that the
prescription of \cite{KLMS} in the context of the decay of unstable
D-brane yielded fortuitously correct result in that the extra pole
contribution turns out absent and that the prescription of
\cite{KLMS} cannot be taken over to other contexts such as the
rolling of accelerated D-brane as was done, for example, in
\cite{OR}.

This paper is organized as follows. In section 2, we first
recapitulate computation of spectral amplitudes and observables for
the decay of unstable D-brane in open and closed string channels, in
flat spacetime, in linear dilaton background, and in two-dimensional
string theory background. We demonstrate that the Fourier transform
between open string and closed string channels is fortuitously
convergent. Consequently, only the naive contributions being
present, the unitarity and the channel duality are obeyed trivially.
In section 3, we study the same for the rolling of accelerated
D-brane, where the acceleration is caused by extremal or
non-extremal NS5-brane background. Here, we find that the extra pole
contribution shows up. Consequently, the optical theorem is seen to
follow only if this extra contribution is taken account of. In
section 4, we highlight important steps in the proposed analytic
continuation of the Fourier transform, and tie up loose ends of
various confusion scattered in the previous works \cite{OR, KLMS}.

\section{Decay of Unstable D-brane}
For completeness and for detailed comparison with rolling D-brane
case, we shall first compute closed string emission out of decaying
D-brane in linear dilaton background following closely the method
employed in the appendix of \cite{KLMS}. The dilaton gradient is set
by:
\be \Phi = {1 \over \sqrt{\alpha'}} (Q\, X^0 + {\bf V} \cdot {\bf
X}), \qquad \mbox{where} \qquad Q \equiv \beta - {1 \over \beta} ~
\qquad(\beta \ge 1)~. \label{dilaton}\ee
This puts the critical dimension $D$ for the bosonic string theory
to be
\be 26 = D - 6 Q^2 + 6 {\bf V}^2, \qquad \mbox{so} \qquad c_{\rm
eff} = 6 Q_\beta^2 - 6 {\bf V}^2~, \ee
where $Q_\beta \equiv (\beta + 1/\beta)$. The effective central
charge $c_{\rm eff}$ sets the growth of density of closed string
states \cite{KutS}:
\begin{eqnarray}
\rho^{(c)}(M) \sim e^{4\pi \sqrt{\frac{c_{\msc{eff}}}{24}
\alpha' M^2}}
\label{dos}
\end{eqnarray}
up to subleading pre-exponential factor of $M$. It grows slower than
the density of states for flat spacetime (obtainable by setting
$Q={\bf V} = 0$).

\subsection{closed string emission}
Consider the decay of an unstable D-brane in linear dilaton
background.
The radiative transition of a D$p$-brane to a single closed string
state of mass $M$ (set by the integer-valued oscillator level
$N=\widetilde{N}$), whose on-shell energy-momentum $(\omega, {\bf
k})$ is given by
\be \Big(\omega_E - {i Q \over \sqrt{\alpha'}}\Big)^2 - \Big({\bf
k}_E + {i {\bf V} \over \sqrt{\alpha'}} \Big)^2 =(\omega^2 -{\bf
k}^2) =  M^2 \quad \mbox{where} \quad \frac{1}{4}\, \alpha' M^2 = N
- {c_{\rm eff} \over 24}, \ee
where $(\omega_E, {\bf k}_{E})$ and $(\omega, {\bf k})$ are
energy-momenta in the Einstein and the string frame, respectively.
In string loop perturbation theory, the transition amplitude is
computed by the disk one-point function $\langle \exp ((-i \omega +
\frac{Q}{\sqrt{\al'}}) X^0) \, \exp ((i {\bf k} + \frac{{\bf
V}}{\sqrt{\al'}}) \cdot {\bf X}) \rangle_{\msc{disk}}$ with the
D$p$-brane boundary condition,\footnote{We only consider the case
when the D-brane has Neumann boundary condition in the space-like
linear dilaton direction.} where the vertex operator is separated
into temporal and spatial parts as indicated. The two parts are
factorized in the gauge that no oscillator in temporal direction is
allowed. Consequently, the transition probability ${\cal P}(\omega)$
of the radiative process is governed entirely by the temporal part
(see (3.29) in \cite{KLMS}):
\be {\cal P}(\omega) = \Big| \left\langle e^{ (- i \omega
+\frac{Q}{\sqrt{\al'}}) X^0} e^{(i {\bf k} + \frac{{\bf
V}}{\sqrt{\al'}})\cdot{\bf X}} \right\rangle_{\msc{disk}} \Big|^2
&=& \Big| {1 \over \beta} \Gamma(1 + i \omega\sqrt{\alpha'} \beta)
\Gamma(- i \omega \sqrt{\alpha'} /\beta)\Big|^2 \nn
&=&{\pi^2/\beta^2 \over \sinh (\pi \omega\sqrt{\alpha'}  \beta)
\sinh (\pi \omega\sqrt{\alpha'}  / \beta)}~. \ee
Then, at leading order in string perturbation theory, the total
number of emitted closed strings from the decay of a D$p$-brane
($p\ge 1$) extended along ${\bf V}$-direction is computed as
\be \overline{\cal N} = N_p^2 V_p\sum_M \sqrt{\rho^{(c)}(M)}
\int_{-\infty}^\infty \!{\rmd^{D-1-p} {\bf k} \over (2 \pi)^{D-1-p}}
\, {1 \over 2 \omega} {\cal P}(\omega)~, \ee \label{closed}
\noindent where the overall coefficient abbreviates $N_p =
\pi^{\frac{D-4}{4}} (2 \pi)^{\frac{D-2}{4}-p}$ and $V_p$ is the
D$p$-brane volume. In (\ref{closed}), the sum is over all final
closed string states of mass $M$ and of oscillator excitations
symmetric between left- and right-moving sectors. Such oscillator
excitations are equivalent in combinatorics to open string
excitation, so the density of the final states is given by
square-root of \eqn{dos}.

Attributed to the Hagedorn growth of the density of states
$\rho^{(c)}(M)$, the total emission number $\overline{\cal N}$ in
(\ref{closed}) (or higher spectral moment) is ultraviolet convergent
so long as linear dilaton has a nonzero spatial component, ${\bf V}
\ne 0$, first observed in \cite{KLMS}. Notice also that temporal
component of the linear dilaton does not alter the ultraviolet
behavior. This is most readily seen for small ${\bf V}$ by expanding
the density of states. To study anatomy of the ultraviolet behavior,
we shall now perform Fourier transformation and re-express
$\overline{\cal N}$ in the open string channel.

\subsection{open string channel viewpoint}

Physical observables such as $\overline{\cal N}$ ought to be
well-defined under the Fourier transform from the closed string
channel to the open string one because
\begin{enumerate}
 \item We start with defining expression of $\overline{\cal N}$,
 consistent with the optical theorem in the closed string channel.
 \item The expression is closed in the Euclidean signature.
Hence we are free from any subtlety that may arise from analytic
continuations between Euclidean and Lorentzian signature of the
spacetime.
\end{enumerate}
As in \cite{KLMS}, expand the transition probability ${\cal
P}(\omega)$ in convergent power series, whose terms are
interpretable as D-instantons arrayed along imaginary time
coordinate:
\be {\cal P}(\omega) = {4 \pi^2\over \beta^2} \sum_{n,m=0}^\infty
e^{-2 \pi \alpha' \omega W(m,n)} \ee
where the location of the D-instantons is denoted as
\be \alpha'  W(m,n) = \sqrt{\alpha'}\Big[\Big(n+{1 \over 2} \Big)
\beta + \Big(m+{1 \over 2} \Big){1 \over \beta}\Big] \ge
\sqrt{\alpha'}~. \ee
So, we take
\be \overline{\cal N} = \Big({2 \pi N_p\over \beta}\Big)^2 V_p
\sum_M \int_{-\infty}^\infty {\rmd^{D-1-p} {\bf k} \over
(2\pi)^{D-1-p}} \sum_{m,n=0}^\infty {1 \over 2 \omega} e^{ - 2 \pi
\alpha' \omega W(m,n)} \ee
and rewrite each D-instanton contribution parametrically via the
closed string channel modulus $t_c$ as
%
%
\be {1 \over 2 \omega} e^{ - 2 \pi \alpha' \omega W(m,n)} &=&
%
\frac{\pi \alpha'}{2} \int_{-\infty}^\infty {\rmd k_0 \over 2 \pi}
\int_0^\infty \rmd {t_c} \, e^{-2\pi t_c \cdot
\frac{1}{4}\alpha'(k_0^2 + {\bf k}^2 + M^2)} e^{2 \pi i \alpha' k_0
W(m,n)}. \ee
This gives
\be \overline{\cal N} \!\!\! &=& \!\!\!\! \Big({2 \pi N_p\over
\beta}\Big)^2 V_p \frac{\pi \alpha'}{2} \sum_{m,n=0}^\infty
\int_0^\infty \rmd {t_c} \int_{-\infty}^\infty {\rmd k_0 \over 2
\pi} \int_{-\infty}^\infty {\rmd^{D-1-p} {\bf k} \over (2
\pi)^{D-1-p}} \, e^{-2\pi t_c \cdot \frac{1}{4}\alpha'(k_0^2 + {\bf
k}^2)} e^{2 \pi i \alpha' k_0 W(m,n)}\nn
&& \hskip3.7cm \times \sum_M \sqrt{\rho^{(c)}(M)} e^{-2\pi t_c \cdot
\frac{1}{4}\alpha' M^2}. \ee
Here, we exchanged order of summations and integrations, and first
performed integrals over off-shell momenta $(k_0, {\bf k})$ and sum
over mass level $M$. The sum over $M$ yields modular covariant
partition function $Z^{(c)}(q_c)$ in terms of the Dedekind eta
function:
\be Z^{(c)} (q_c) &\equiv& \sum_M \sqrt{\rho^{(c)}(M)} \, \,
q_c^{\frac{1}{4}\alpha'M^2} \qquad \mbox{where} \qquad q_c \equiv
e^{-2\pi t_c} \nn &=& \eta^{-(D-2)}(q_c)~. \ee
Integrations over the $(D-p)$-dimensional momenta $(k_0, {\bf k})$ yield
$(2 \pi^4 \alpha' t_c)^{-(D-p)/2}$ times Gaussian damping factor
$e^{-2\pi \alpha' W^2(m,n)/ t_c}$. We now perform modular
transformation to the open string channel $t_c = 1/t_o$, where $t_o$
is modulus of the open string channel and $q_o \equiv e^{- 2\pi
t_o}$. Putting all these together, we finally have
\be \overline{\cal N} =  C_p \, V_p
 \sum_{m,n=0}^\infty \int_0^\infty {\rmd t_o \over t_o} t_o^{-p/2}\, e^{- 2\pi
t_o \alpha' W^2(m,n) } \, \eta^{-(D-2)} (q_o), \label{openexp}\ee
with $C_p = \Big({2 \pi N_p\over \beta}\Big)^2 \frac{\pi
\alpha'}{2}(2\alpha'\pi^4)^{-\frac{D-p}{2}}$, reproducing the result
reported in \cite{KLMS}. As it stands, the final expression
\eqn{openexp} is at odd to the intuition based on, for example, the
Schwinger pair production in (time-dependent) electric field, since
the integral over the open string modulus $t_o$ is still intact. If
the total emission number is interpretable as arising from on-shell
two-particle branch cut in the open string channel, the modulus
integral ought to be absent! Therefore, To understand underlying
physics better, we shall now compute the cylinder amplitude directly
and then extract the imaginary part via the optical theorem.

\subsection{Lorentzian cylinder amplitude}
Unitarity and optical theorem thereof, combined with the open-closed
string channel duality, should enable us to extract the emission
number $\overline{\cal N}$ of closed strings from decaying D$p$-brane
as the imaginary part of the cylinder amplitude. In the closed
string channel diagram, the computation reduces to \eqn{closed}, as
in quantum field theory. It is, however, somewhat nontrivial to
evaluate the imaginary part of the cylinder amplitude directly from
the open string channel. Here we present the {\sl ab initio}
derivation, refining that in the text of \cite{KLMS}, by starting
with manifestly well-defined Lorentzian cylinder amplitude.

We begin with the cylinder amplitude in the closed string channel in
which both the worldsheet and the target spacetime signatures are
taken Lorentzian. Omitting overall numerical factors for the moment,
the amplitude is given by
\begin{eqnarray}
Z_{\rm cylinder} = i \pi \alpha' V_p \int_{s_c^{\rm IR}}^{s_c^{\rm
UV}} \rmd s_c \int_{-\infty}^{\infty} \frac{\rmd \omega_L}{2\pi} \,
\frac{\pi^2/\beta^2 \cdot
q_c^{-(1-i\hat{\epsilon})^2\frac{1}{4}\alpha'\omega_L^2}
}{\sinh(\pi\beta\omega_L \sqrt{\alpha'}) \sinh(\pi\omega_L
\sqrt{\alpha'}/\beta)} \, Z_{\cM}^{(c)}(q_c) \ , \label{cl}
\end{eqnarray}
where $q_c = e^{2\pi i \tau_c}$ with $\tau_c = s_c + i\epsilon$, and
$Z_{\cM}(q_c)$ represents the contribution from the closed string
zero-modes and oscillator parts~\footnote{We are using different
normalization for modulus parameters from \cite{KLMS}: $t$(KLMS) =
$(\pi/4)t$(here). In addition, they adopted $\alpha'=1$
convention.}. The Lorentzian worldsheet is regularized by $i
\epsilon$ prescription, while the Lorentzian spacetime is
regularized by $-i \hat{\epsilon}$-prescription. $s_c^{\rm {UV}}$
($s_c^{{\rm IR}}$) is an ultraviolet (infrared) regulator of the closed
string channel modulus. With these prescriptions, the integral over
$\om_L$ is convergent so long as $2 \hat{\ep} s_c^{\rm UV} > \ep >0$
is retained.

Defining the open string modular parameter
as $q_o = e^{-2\pi i \tau_o}$ where
$\tau_o = s_o - i \epsilon$ with $s_o = 1/s_c$, one can rewrite
\eqref{cl} in terms of open string channel energy $\omega_L'$ as
\begin{eqnarray}
Z_{\rm cylinder} &=&  V_p \int_{s_o^{\rm UV}}^{s_o^{\rm IR}} {\rmd
s_o} \int_{-\infty}^{\infty} \rmd \omega'_L \left(i \pi \alpha'
\int_{-\infty}^{\infty} \rmd \omega_L \frac{\cos(\pi \alpha'
\omega_L \omega'_L)}{\sinh(\pi\beta \omega_L \sqrt{\alpha'} )
\sinh(\pi\omega_L\sqrt{\alpha'}/\beta)} \right) \nonumber \\
&& \hskip4cm \times \,
q_o^{-(1+i\hat{\epsilon}')^2\frac{1}{4}\alpha'{\omega_L'}^2}
Z_{\cM}^{(o)}(q_o) \, ,
\end{eqnarray}
where $s^{\rm IR}_o\equiv 1/s^{\rm UV}_c$, $s^{\rm UV}_o\equiv
1/s^{\rm IR}_c$ are the cut-off's in the open string modulus.
As opposed to the closed string channel, we have to adopt the
$+i\hat{\ep}'$-prescription for the Lorentzian space-time, and
the above integral is well-defined as long as
$2 \hat{\ep}' s^{\rm UV}_o > \ep$.
The expression in the large parenthesis yields the open string
density of states, $\rho^{(o)}(\omega'_L)$. It is infrared divergent
at $\omega_L= 0$. To regularize it, we subtract minimally the double
pole~\footnote{This subtraction does not affect the imaginary part
of the partition function we are primarily interested in.} so that
\begin{eqnarray}
\rho^{(o)}(\omega'_L)_{\mathrm{reg}} &=& i \pi \alpha'
\int_{-\infty}^{\infty} \rmd \omega_L \left(\frac{\cos(\pi\alpha'
\omega_L\omega'_L)}{\sinh(\pi\beta\omega_L \sqrt{\alpha'})
\sinh(\pi\omega_L \sqrt{\alpha'}/\beta)} - \frac{1}{\pi^2
\alpha'\omega_L^2} \right) \cr &=& -{2}\partial_{\omega'_L} \log
S_\beta\left({Q}_\beta + i\sqrt{\alpha'}\omega'_L\right) \ , \label{dens}
\end{eqnarray}
where the `$q$-Gamma function' $S_\beta(x)$ is defined by~\footnote
  {Here the normalization of variable $x$ differs with factor 2
   from the one given in \cite{FZZ}.}
\begin{eqnarray}
-\partial_x \log S_\beta (x) = \int_{-\infty}^{\infty} \rmd t
\left(\frac{\cosh((x-Q_\beta)t)}{2\sinh(\beta t)\sinh(t/\beta)}
-\frac{1}{2t^2} \right)\
\end{eqnarray}
for $\mathrm{Re}(x) < 2 Q_\beta$ and analytically continued to the
whole complex plane~\footnote{Notice that the Lorentzian density
\eqref{dens} is well-defined without the analytic continuation. We
stress that this should be contrasted against the approach of
\cite{KLMS}.}. See, for example, \cite{FZZ,Nakayama-review}.

Now we perform the Wick-rotation both in the target space and on the
worldsheet. First, Wick rotate the open string channel energy as
$\om'_L\, \rightarrow\, e^{i(\frac{\pi}{2}-0)} \om'_L$ and set
$\om'_L = i \om'$ $(\om' \in \br)$. Then, we can safely Wick rotate
the worldsheet Schwinger parameter as $s_o \, \rightarrow \, - i
t_o$ ($t>0$). Notice that we will need to perform the Euclidean
rotation in opposite direction for the closed and the open string
channels due to the difference of the $i\ep$-prescription.
 There is no obstruction in such contour deformation because
 $\partial_x \log S_\beta (x) $ has poles only on the real axis.
 We will see that this is specific to the decaying D-brane situation
 and do not hold generally. In fact, in section 3
 dealing with the rolling D-branes,
 we shall show that there exist extra contributions from crossing poles
 in the course of the contour rotation and that their contributions are
 essential for maintaining the unitarity.
 After Wick rotating the worldsheet,
the cylinder amplitude in the open string sector
 is given by
\begin{eqnarray}
Z_{\rm cylinder} = -2 V_p \int_{0}^\infty \rmd t_o
\int_{(1-i0)\mathbb{R}} \rmd \omega' \partial_{\omega'} \log S_\beta
\left({Q_\beta} -\sqrt{\alpha'}
\omega'\right)q_o^{\frac{1}{4}\alpha'{\omega'}^2} Z_M^{(o)}(q_o) \ .
\end{eqnarray}

Imaginary part of the partition function comes from the simple poles
of the $q$-Gamma function $S_\beta \left(Q_\beta -\sqrt{\alpha'}
\omega'\right)$ at $\frac{1}{2}\omega' = W(m,n)$
for $n,m \in \mathbb{Z}_{\ge 0} $ and simple zeros for $n,m \in
\mathbb{Z}_{< 0}$. Therefore, collecting imaginary parts from the
contour integration over $\omega'$ and applying the optical theorem,
we finally obtain
\begin{eqnarray}
\overline{\cal N} =  \mathrm{Im}\, Z_{\rm cylinder} = C_p \, V_p
\sum_{n,m=0}^\infty \int_0^\infty {\rmd t_o \over t_o}
t_o^{-\frac{p}{2}}\, e^{-2\pi t_o \alpha' W^2 (m,n)} \,
 \eta^{-(D-2)}(q_o) \ ,
\end{eqnarray}
where we have evaluated the free oscillator part explicitly and
reinstated overall numerical factors.
This is in perfect agreement with \eqref{dilaton}, and it may be
interpreted as a nontrivial check of unitarity and open-closed
duality in the Lorentzian signature.

\subsection{D-brane decay in two-dimensional string theory}
In a similar method, one can compute the spectral observables from
the D-brane decay in two-dimensional string theory. The boundary
state for the unstable D-brane in two-dimension is given by the
ZZ-brane boundary state \cite{ZZ}:
\begin{eqnarray}
\langle e^{(i k + 2/\sqrt{\alpha'})\phi} \rangle_{\msc{disk}} =
\mu^{-\frac{i}{2} \sqrt{\alpha'} k} \frac{2
\sqrt{\pi}}{\Gamma(1-ik\sqrt{\alpha'})\Gamma(ik\sqrt{\alpha'})} \ .
\end{eqnarray}
Combining it with the rolling tachyon boundary states, the total
emission number of closed string is given by
\begin{eqnarray}
\overline{\cal N} = N_o^2 \int^{\infty}_{0} \rmd k \int^{\infty}_{0}
\frac{\rmd \omega}{2\omega}  {\cal P} (\omega, k) \delta(\omega-k) \
,
\end{eqnarray}
where the on-shell condition $\omega = k$ is imposed, and the
transition probability is
\begin{eqnarray}
{\cal P} (\omega, k)  = \left| \langle e^{- i\omega X^{0}} e^{
(ik+2/\sqrt{\alpha'})\phi} \rangle_{\msc{disk}} \right|^2 =
\frac{\sinh^2(\pi k\sqrt{\alpha'})}{\sinh^2(\pi \omega
\sqrt{\alpha'})} \ .
\end{eqnarray}
We see that, after performing the $k$-integration, the resultant
total emission number is ultraviolet divergent.

To express $\overline{\cal N}$ in open string channel, we repeat the
analysis of section 2.2 and expand the transition probability in
arrays of imaginary D-instantons. The result is
\begin{eqnarray}
\overline{\cal N} &=& N_o^2 \sum_{m,n = 0}^\infty \int_0^\infty \rmd
k \int_0^\infty \rmd t_c \int_{-\infty}^{\infty} \frac{\rmd
k_0}{2\pi} e^{-2\pi t_c \cdot \frac{1}{4}\alpha' (k_0^2 + k^2)}
e^{2\pi i \alpha' k_0 W(m,n)} \sinh^2(\pi k\sqrt{\alpha'}) \Big\vert_{\beta \rightarrow 1} \cr
&=& N_o^2 \sum_{m,n = 0}^\infty \int_{0}^\infty \frac{\rmd t_o}{t_o}
\Big({1 \over q_o} - 1 \Big) q_o^{ \alpha' W^2(m,n)}
\Big\vert_{\beta \rightarrow 1}
 \ , \label{twodim}
\end{eqnarray}
where we have reinstated $W(m,n)$ for the purpose of
regularization~\footnote{Because of the subtraction of singular
vector in $(1/q_o - 1)$, the resultant amplitude is {\it
non-unitary}.}. The expressoin exhibits ultraviolet divergence as
$t_o \to \infty$.

On the other hand, it is possible to obtain the same radiation rate
from the direct evaluation of the imaginary part of the Lorentzian
cylinder amplitude in the open-string channel as was done in section
2-3:
\begin{eqnarray}
Z_{\rm cylinder} = i N_o^2 \int_0^\infty \rmd s_c
\int_{-\infty}^{\infty} \frac{\rmd \omega_L}{2\pi} \int_0^{\infty}
\frac{\rmd k}{2\pi} \, \frac{\sinh(\pi \sqrt{\alpha'} k)^2
}{\sinh(\pi \sqrt{\alpha'} \omega_L)^2}\,
q_c^{\frac{1}{4}\alpha'(-\omega^2_L + k^2)}  \ .
\end{eqnarray}
After rewriting the open string density by the $q$-Gamma function as
in section 2.3, we obtain open string channel expression of the
partition function. We then find the imaginary part from the poles
located at $\frac{1}{2}\omega' = W(m,n)$, and reproduce
\eqref{twodim}. This confirms that the partition function is
manifestly unitary, obeying the optical theorem. Here again, the
regularization $\beta \to 1$ is implicit.

\section{Rolling  of Accelerated D-brane}

Consider next rolling dynamics of accelerated D-brane. We first
recapitulate rolling D-brane dynamics in the fivebrane background.
For details, we refer the readers to
\cite{rolling,NST,NRS}. The fermionic string
background of interest is described by the exact conformal field
theory:
\be \br_t \times \br_{\phi} \times \cM \qquad \mbox{or} \qquad
{\mbox{SL}(2)_k \over \mbox{U}(1)} \times \cM \ ,
\label{background}\ee
where $\br_t$ is the timelike free theory, $\br_{\phi}$ is the
spacelike linear dilaton with slope $Q \equiv \sqrt{2/k}$, and $\cM
$ is a unitary rational conformal field theory describing the
transverse geometry. The conformality condition relates the dilaton
slope $Q$~\footnote
{Here we take a different convention of the dilaton background from
the previous section and set $\Phi(\phi) =
\frac{Q}{\sqrt{2\al'}}\phi$. The contribution to the central charge
of the linear dilaton background is now $\Delta c = 3 Q^2$, whereas
that of the previous section is $\Delta c= 6Q^2$. }
to the central charge of ${\cal M}$:
\begin{eqnarray}
 c_{\cM}+ (3 + 3 Q^2) = 15~.
\end{eqnarray}
We shall focus on the region $Q \le \sqrt{2}$ viz. $k \ge 1$
for the moment, which corresponds to the `black-hole phase' of the
super-coset SL(2)$_k$/U(1) conformal field theory \cite{GKRS}.
In the background of fivebranes, $k = 1, 2, \cdots $
counts the total number of fivebranes \cite{NS5brane,CHS}.

Notice that the effective central charge of the background
\eqn{background} is equal $c_{\cM}$;
\begin{eqnarray}
 && c_{\msc{eff}} \equiv 12 -24 \cdot \frac{Q^2}{8}
= 12\Big(1-{1 \over 4}
 Q^2\Big)
 = c_{\cM}~.
\end{eqnarray}
Notice also that, from Cardy's formula, the density of closed string
state grows at ultraviolet asymptotically as
\begin{eqnarray}
 && \rho^{(c)}(M) \sim
e^{4\pi \sqrt{\frac{c_{\msc{eff}}}{12}\cdot \frac{1}{2} \alpha'
M^2}} = e^{4\pi \sqrt{\left(1-\frac{1}{4}Q^2\right)\cdot
\frac{1}{2}\alpha' M^2}}
\end{eqnarray}
up to pre-exponential factors of $M$.

\subsection{closed string emission}
Consider a D0-brane placed initially in the background of
\eqn{background}. Subsequently, the D0-brane rolls toward
the five-branes and forms a non-threshold bound-state.
The emission
number $\overline{\cal N}$ of the process is determined by the
boundary wave function $\Psi(p, \omega)$ of the rolling D0-brane and
was computed in \cite{NST,Sahakyan,NPRT} for
the extremal NS5-branes.
Explicitly, for the
NS-sector,
\be \overline{\cal N} = N^2_{\rm NS} \sum_M \sqrt{\rho^{(c)}(M)}
\int_0^{\infty} {\rmd p \over 2 \pi} \, \frac{1}{2\om} {\cal P}(p,
\omega) ~, \label{ImZ 0}\ee
where $N_{\rm NS}$ is an appropriate numerical factor normalizing
the boundary states, $\om = \sqrt{p^2 + M^2}$ is the on-shell energy
of the emitted closed string, and
\be {\cal P}(p, \omega) \equiv \left|\Psi(p,\om)\right|^2
 = \frac
{\sinh\frac{2\pi}{Q}\sqrt{\frac{\alpha'}{2}} p} {\left(\cosh
\frac{2\pi}{Q} \sqrt{\frac{\alpha'}{2}} p + \cosh \frac{2\pi}{Q}
\sqrt{\frac{\alpha'}{2}} \om \right) \sinh \pi Q
\sqrt{\frac{\alpha'}{2}}  p }~ \label{powerspec}
\end{eqnarray}
is the transition probability.
Essentially the same formula is also applicable to the `incoming
radiation' of rolling D-brane in the non-extremal NS5-brane
background \cite{NRS}.
Notice that ${\cal P}(p, \omega)$ is
a function of both $\omega$ and $p$, whereas it depended only on
$\omega$ for the decaying D-branes considered in the previous
section. This is because the unstable D-brane stays at rest during
the decay process, while here the accelerated D-brane rolls along
the direction $\mathbb{R}_\phi$.

In \cite{NST,NPRT,NRS}, the total emission number $\overline{\cal
N}$ was computed in the closed string channel via the optical
theorem $\overline{\cal N} = \mbox{Im}\, Z_{\msc{cylinder}}$, where
$Z_{\msc{cylinder}}$ is the cylinder amplitude with rolling D-branes
and the imaginary part arose from on-shell closed string states.

\subsection{open string channel viewpoint}
What is the nature of the ultraviolet behavior of the emission
number $\overline{\cal N}$ and how is it compared to the decay of
rolling D-brane? To answer these, we shall now recast \eqn{ImZ 0} in
the open string channel, following technical procedures considered
in the previous section and appendix of \cite{KLMS}.

We begin with expanding the transition probability ${\cal P}(p,\om)$
of the D0-brane \eqn{powerspec} in power series of contribution of
imaginary branes:
\begin{eqnarray}
 && {\cal P}(p,\om) = \sum_{n=1}^{\infty}
a_n(p) e^{-2\pi n\frac{\om}{Q}\sqrt{\frac{\alpha'}{2}}} ~,
\label{expansion}
\\
 && a_n(p)= 2(-1)^{n+1}
\frac{\sinh\left(\frac{2\pi n}{Q} \sqrt{\frac{\alpha'}{2}} p \right)} {\sinh
(\pi Q \sqrt{\frac{\alpha'}{2}} p)}~. \label{a n}
\end{eqnarray}
%
%
As before,
we parametrically rewrite \eqn{ImZ 0} as
\begin{eqnarray}
\overline{\cal N} &=&
 N_{\rm NS}^2 \sum_M \sqrt{\rho^{(c)}(M)} \nn
 &\times& \int_0^{\infty} {\rmd p
\over 2
 \pi} \sum_{n=1}^{\infty} \int_{-\infty}^{\infty}
{\rmd k_0 \over 2 \pi}\,
 \int_0^{\infty} \frac{\alpha'}{2} \rmd t_c \,
  a_n(p) e^{\frac{2\pi i n}{Q}\sqrt{\frac{\alpha'}{2}} k_0}
 e^{-2\pi t_c \frac{1}{4}\alpha'\left(k_0^2+p^2+M^2\right)}~, \qquad
\label{ImZ 1}
\end{eqnarray}
by introducing the Schwinger parameter $t_c$
in the closed string channel~\footnote
  {Strictly speaking, we could have the closed string tachyon
   $M^2<0$, and the rewriting \eqn{ImZ 1} would not be
   completely correct due to the infrared divergence. We can
   avoid this difficulty by considering the GSO projected
   amplitude. We are concerned with the large $M$ asymptotics,
   so shall go on ignoring it to avoid unessential complexity.
   }.

We now evaluate each contribution separately. Begin with the sum
over the transverse mass $M$. By definition, the sum gives modular
invariant cylinder amplitude of the $\cM$-sector:
\begin{eqnarray}
\sum_M \sqrt{\rho^{(c)}(M)} e^{-2\pi t_c \frac{\alpha'}{4} M^2} &=&
Z^{(c)}_{\cM}(q_c) \qquad \mbox{where} \qquad q_c = e^{- 2 \pi t_c}
\nn &=& Z^{(o)}_{\cM}(q_o) \qquad \mbox{where} \qquad q_o = e^{- 2
\pi t_o} \quad (t_o \equiv 1/t_c)
\end{eqnarray}
by applying the standard open-closed duality and expressing the
result in terms of the open string Schwinger parameter $t_o$.

The amplitude $Z^{(o)}_{\cM}(t_o)$ asymptotes at large $t$ to
(corresponding to the ultraviolet behavior in the closed string
channel):
\begin{eqnarray}
 Z_{\cM}^{(o)}(t_o) \sim t_o^{\gamma}\,
e^{2\pi t_o \cdot \frac{c_{\cM}}{24}}
 = t_o^{\gamma} \, e^{\pi t_o \left(1-\frac{Q^2}{4}\right)} \qquad
 \mbox{for} \qquad t_o \, \rightarrow\, +\infty.
\end{eqnarray}
Here, the exponent $\gamma$ is determined by the number of
non-compact Neumann directions
in the $\cM$-sector. Such details,
however, are not relevant for our discussions.

The Gaussian integral over $k_0$ is readily evaluated,
resulting in
\begin{eqnarray}
 && \overline{\cal N}
= N^2_{\rm NS} \sqrt{\frac{\alpha'}{2}} \int_0^{\infty} {\rmd p \over 2 \pi}
\sum_{n=1}^{\infty}
  \int_0^{\infty} \frac{\rmd t_o}{t_o^2} \sqrt{t_o} \,
   a_n(p) \, e^{-\pi t_o \frac{n^2}{Q^2} -\frac{2\pi}{t_o} \frac{\alpha'}{4}p^2}
  \cdot Z^{(o)}_{\cM}(t_o)~.
\end{eqnarray}
The $k_0$-integral yields the Boltzmann factor with the temperature
determined by the Euclidean periodicity
($1/Q$ in our case) for the `hairpin brane'
\cite{hairpin}, which is the Euclidean rotation of the rolling
D-brane, as clarified in \cite{NRS,KutAcc}.
This is essentially the same as the standard argument
for thermal tachyon in the thermal string theory
\cite{thermal closed}.

Our goal is to re-express the rate \eqn{ImZ 0} in the open string
channel, so we shall Fourier transform the closed string momentum
$p$ to the open string momentum $p'$. This requires a careful
treatment, because the momentum-dependent coefficients $a_n(p)$ in
\eqn{a n} could be exponentially growing functions. In such cases,
the Fourier transform may not exists in a naive sense. We start with
the identity:
\begin{eqnarray}
 && e^{-2\pi t_c \cdot \frac{1}{4} \alpha' p^2} = \sqrt{t_o} \int_{\mathbb{R}+i\xi}
 \sqrt{\frac{\alpha'}{2}}\rmd p'\,
e^{-2\pi t_o \cdot \frac{1}{4} \alpha' p^{'2}+2\pi i \cdot
\frac{1}{2} \alpha' p p'} \qquad \mbox{for} \qquad \xi \in \br~.
\label{gauss xi}
\end{eqnarray}
In the $p$-integral, the function $e^{2\pi i \frac{1}{2} \alpha' p
p'}$ works as a damping factor and renders the integral finite if
the parameter $\xi$ is chosen suitably. For later convenience, we
shall decompose $a_n(p)$ as
\begin{eqnarray}
 && a_n(p) = a_n^+(p)-a_n^-(p)~, \nn
 && a^{\pm}_n(p) \equiv (-1)^{n+1}
\frac{e^{\pm\frac{2\pi n}{Q} \sqrt{\frac{\alpha'}{2}} p}}
   {\sinh (\pi Q p \sqrt{\frac{\alpha'}{2}})}~.
\label{a pm}
\end{eqnarray}
%
Observing the asymptotic behavior of the coefficients $a^+_n(p)$, we
readily find that the closed string channel momentum integral $\dsp
\int \rmd p\, a^+_n(p) \, e^{2\pi i \cdot \frac{1}{2} \alpha' p p'}$
is well-defined as long as $\xi^+_n$ is chosen within the range
$(\frac{n}{Q}-\frac{Q}{2})<\sqrt{\frac{\alpha'}{2}}\xi^+_n <
(\frac{n}{Q} + \frac{Q}{2})$. We can then safely exchange the order
of the integrals.
Carrying out the $p$-integral first, we find~\footnote
  {Here, we are temporarily shifting
  the contour as $\br \, \rightarrow\, \br-i0$
  to avoid the pole $p=0$.
   We eventually restore it back to $\br$ {\sl after}
  taking the difference
  $a_n(p) \equiv a_n^+(p)-a_n^-(p)$.
  The final result \eqn{ImZ final} remains intact,
  even if another contour shift $\br+i0$ is taken, as is
  easily checked.}
\begin{eqnarray}
  && \int_{\mathbb{R}-i0}\rmd p\, a^+_n(p)
  e^{-2\pi t_c \frac{\alpha'}{4} p^2}
= (-1)^{n+1}\frac{i\sqrt{t}}{Q} \int_{\mathbb{R} + i\xi^+_n} \rmd
p'\, \frac {e^{\pi \left(\sqrt{\frac{\alpha'}{2}}\frac{p'}{Q}-i
\frac{n}{Q^2}\right) -2\pi t_o \frac{\alpha'}{4} p^{'2}}} {\cosh \pi
\left(\sqrt{\frac{\alpha'}{2}}\frac{p'}{Q}-i\frac{n}{Q^2}\right)}~.
\qquad  \label{evaluation 2}
\end{eqnarray}
Finally, we shift the contour back:
$\br+i\xi_n^+ \, \rightarrow\,
\br$ so that  the open string momentum $p'$ is real-valued. In this
step, we cross the poles so need to take care of pole contributions.
(See Figure 1.)
\vskip0.5cm
\begin{figure}[htbp]
    \begin{center}
    \includegraphics[width=0.5\linewidth,keepaspectratio,clip]
      {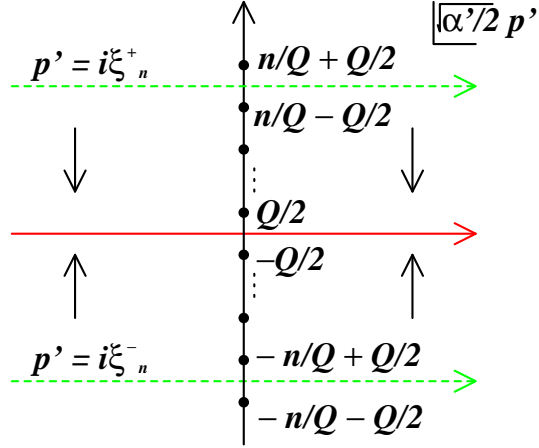}
    \end{center}
    \caption{\sl Deformation of the contour from the broken line
    to the solid line picks up pole contributions.}
    \label{contour1}
\end{figure}
\vskip0.5cm

The relevant poles are located at
\begin{eqnarray}
 && \sqrt{\frac{\alpha'}{2}}p' = i {\al_m }~, ~~~ \al_m \equiv \frac{n}{Q}
-Q\left(m+\frac{1}{2}\right) \quad \mbox{where} \quad m=0,1, \ldots,
\left\lb \frac{n}{Q^2}-\frac{1}{2}\right\rb~, \qquad
\quad\label{poles}
\end{eqnarray}
where $\lb ~~ \rb$ denotes the Gauss symbol, and their residues are
evaluated as $(-1)^{n+1}\frac{i}{\pi} e^{\pi t_o \al_m^2}$.
We thus obtain
\begin{eqnarray}
   \int_{\mathbb{R}-i0}\sqrt{\frac{\alpha'}{2}}\rmd p\, a^+_n(p)
   e^{-2\pi t_c \frac{\alpha'}{4} p^2}
&=&  (-1)^{n+1}\frac{i\sqrt{t_o}}{Q} \int_{-\infty}^{\infty}
\sqrt{\frac{\alpha'}{2}}\rmd p'\, \frac {e^{\pi
\left(\sqrt{\frac{\alpha'}{2}}\frac{p'}{Q}-i \frac{n}{Q^2}\right)
-2\pi t_o \frac{\alpha'}{4} p^{'2}}} {\cosh \pi
\left(\sqrt{\frac{\alpha'}{2}}\frac{p'}{Q}-i\frac{n}{Q^2}\right)}
\nn &+& 2(-1)^{n+1} \sqrt{t_o} \sum_{m=0}^{\left\lb
\frac{n}{Q^2}-\frac{1}{2}\right\rb}\, e^{\pi t_o \left[
\frac{n}{Q}-Q\left(m+\frac{1}{2}\right) \right]^2}~.
\label{evaluation 3}
\end{eqnarray}
The integral of $a^-_n(p)$ is calculated in a similar
way. This time, we should start with the contour $\br+i \xi^-_n$
with $(-\frac{n}{Q}- \frac{Q}{2}) < \sqrt{\frac{\alpha'}{2}}\xi^-_n <
(-\frac{n}{Q}+\frac{Q}{2})$ and, after performing the $p$-integral
first, again shift it back to $\br+i\xi^-_n\,\rightarrow\,\br$. The
relevant pole contributions come from $\sqrt{\frac{\alpha'}{2}}p'=-i\al_m$
($m=0,1,\ldots, \left\lb \frac{n}{Q^2}-\frac{1}{2}\right\rb$), and
we obtain
\begin{eqnarray}
\int_{\mathbb{R}-i0} \sqrt{\frac{\alpha'}{2}} \rmd p\, a^-_n(p)\,
    e^{-\pi t_c \alpha' p^2}
&=&  (-1)^{n+1}\frac{i\sqrt{t_o}}{Q} \int_{-\infty}^{\infty}
\sqrt{\frac{\alpha'}{2}}\rmd p'\, \frac {e^{\pi
\left(\sqrt{\frac{\alpha'}{2}}\frac{p'}{Q}+i \frac{n}{Q^2}\right)
-2\pi t_o \frac{\alpha'}{4} p^{'2}}} {\cosh \pi
\left(\sqrt{\frac{\alpha'}{2}}\frac{ p'}{Q}+i\frac{n}{Q^2}\right)}
\nn &-& 2(-1)^{n+1} \sqrt{t_o} \sum_{m=0}^{\left\lb
\frac{n}{Q^2}-\frac{1}{2}\right\rb}\, e^{\pi t_o \left[
\frac{n}{Q}-Q\left(m+\frac{1}{2}\right) \right]^2}~.
\label{evaluation 4}
\end{eqnarray}
Notice that the relative sign change in the pole term compared to
$a^+_n(p)$ integral originates from the orientation
of integration contour surrounding each pole. Therefore,
we find
\begin{eqnarray}
\int_0^{\infty}\sqrt{\frac{\alpha'}{2}}\rmd p\, a_n(p)e^{-2\pi t_c \frac{\alpha'}{4}
p^2} &=& \frac{1}{2} \int_{-\infty}^{\infty} \sqrt{\frac{\alpha'}{2}}\rmd p\,
\left(a_n^+(p) - a_n^-(p)\right)e^{-2\pi \alpha' \frac{t_c}{4} p^2} \nn
&=& (-1)^{n+1}\frac{\sqrt{t_o}}{Q} \int_{-\infty}^{\infty}\sqrt{\frac{\alpha'}{2}}\rmd p'\,
\frac{\sin \left(\frac{2\pi n}{Q^2}\right) e^{-2\pi t_o \frac{\alpha'}{4}
p^{'2}}} {\cosh \left(\frac{2\pi}{Q}\sqrt{\frac{\alpha'}{2}} p'\right) +\cos
\left(\frac{2\pi n}{Q^2}\right)} \nn
&+& 2 (-1)^{n+1} \sqrt{t_o}
\sum_{m=0}^{\left\lb \frac{n}{Q^2}-\frac{1}{2} \right\rb} \, e^{\pi
t_o \left( \frac{n}{Q}-Q\left( m+\frac{1}{2}\right) \right)^2}~.
\label{evaluation 5}
\end{eqnarray}
In this way,
we derive the desired open string channel expression of
the total radiation rate;
\begin{eqnarray}
 \overline{\cN} &=& N^2_{\rm NS} \int_0^{\infty} \frac{\rmd t_o}{t_o}\,
\Big( F_{\rm naive} (t_o) + F_{\rm pole} (t_o) \Big)~, \nn
 F_{\rm naive} (t_o) &=& \frac{1}{Q}\int_{-\infty}^{\infty} \sqrt{\frac{\alpha'}{2}} \rmd p'
\sum_{n=1}^{\infty} \, (-1)^{n+1} \frac{\sin \left(\frac{2\pi
n}{Q^2}\right) e^{-\pi t_o \left( \frac{\alpha'}{2} p^{'2} +
\frac{n^2}{Q^2}\right)}} {\cosh \left(\frac{2\pi}{Q}\sqrt{\frac{\alpha'}{2}}
p' \right) + \cos \left(\frac{2\pi n}{Q^2}\right)} \,
Z_{\cM}^{(o)}(t_o) \nn F_{\rm pole}(t_o) &=& 2 \sum_{n=1}^{\infty}
(-1)^{n+1} \sum_{m=0}^{\left\lb \frac{n}{Q^2}-\frac{1}{2}\right\rb}
\,e^{\pi t_o \left[Q^2\left(m+\frac{1}{2}\right)^2- 2n
\left(m+\frac{1}{2}\right)\right]}\, Z_{\cM}^{(o)}(t_o)~. \label{ImZ
final}
\end{eqnarray}
The first term in \eqn{ImZ final} coincides with the total radiation
claimed by \cite{OR} modulo inessential numerical factor~\footnote
  {It differs slightly from the one
   given in \cite{OR} in that we study
   the fermionic string, while \cite{OR} studies the bosonic string.}.
   It remains finite as $t_o\,\rightarrow\, + \infty$. The second
term, which the analysis of \cite{OR} missed altogether, is of
crucial importance. It is evident that the $m=0$ term is the leading
contribution for each $n$. Recalling $Z^{(o)}_{\cM}(t_o)\, \sim \,
e^{\pi t_o \left(1-\frac{Q^2}{4}\right)}$ asymptotically (up to
pre-exponential power corrections), each $m=0$ term behaves as
\begin{eqnarray}
 \sim e^{\pi t_o \left(\frac{Q^2}{4}- n\right)
+ \pi t_o \left(1-\frac{Q^2}{4}\right)} = e^{\pi t_o (1-n)} \qquad
\mbox{as} \qquad t_o\,\rightarrow\, + \infty~. \label{evaluation m=0
term}
\end{eqnarray}
Therefore, we get the leading contribution from the $n=1$ term,
which shows a massless behavior. Hence, we have reproduced
the Hagedorn-growth behavior expected in \cite{NST,Sahakyan,
NPRT,NRS}. Notice that all the $n>1$
contributions are massive, and thus are not relevant in the
ultraviolet regime of closed string radiations.


\subsection{Lorentzian cylinder amplitude}
In the previous section, we recasted the total emission number
$\overline{\cN}$ of the rolling D0-brane, defined as the sum over
the on-shell states of emitted closed string \eqn{ImZ 0}, in the
open string channel. Now, by the optical theorem and the channel
duality, we ought to be able to obtain $\overline{\cN}$ equally well
from the cylinder amplitude evaluated in the open channel. In this
section, we shall compute explicitly the cylinder amplitude in the
open string channel and show that its imaginary part reproduces
precisely the result \eqn{ImZ final}. This would serve as a
non-trivial check-point of our previous analysis for the consistency
with unitarity and the open-closed channel duality. Notice in
particular that the channel duality is far from being obvious in the
worldsheet in Lorentzian signature. For definiteness, we continue to
focus on the NS sector.

We start with the cylinder amplitude with Lorentzian
worldsheet~\footnote
  {Here, we stress the importance of
   taking the worldsheet Lorentzian.
   The Fourier transformation from the closed to
   open channel is well-defined only for
   the Lorentzian $\om_L$ in spacetime. Accordingly,
    we need to take
   the Lorentzian worldsheet so that the cylinder amplitude
   becomes well-defined.} $Z_{\rm cylinder}$:
%
%
\begin{eqnarray}
&& Z_{\rm cylinder} \nn
&=& i \frac{\alpha'}{2} \int_{s_c^{\rm
UV}}^{s_c^{\rm IR}} \rmd s_c \int_{0}^{\infty}\rmd p
\int_{-\infty}^{\infty} \rmd \om_L \, \frac {\sinh
\left(\frac{2\pi}{Q}\sqrt{\frac{\alpha'}{2}}p\right)}
{\left[\cosh\left(\frac{2\pi}{Q} \sqrt{\frac{\alpha'}{2}}
\om_L\right) +\cosh\left(\frac{2\pi}{Q}\sqrt{\frac{\alpha'}{2}}p
\right)\right] \sinh(\pi Q \sqrt{\frac{\alpha'}{2}} p)} \nn
&& \hskip3cm \times
\frac{q_c^{\frac{1}{4}\alpha'(p^{2}-(1-i\hat{\ep})^2\om_L^{2})}}
{\eta(q_c)^2} \frac{\th_3(q_c)}{\eta(q_c)}
 \cdot Z_{\cM}^{(c)}(q_c) \cdot \eta(q_c)^2
 \frac{\eta(q_c)}{\th_3(q_c)}~.
 \label{zl}
\end{eqnarray}
Here, we again adopt the $i\ep$-prescription for the Lorentzian
worldsheet, while the $-i \hat{\ep}$-prescription for the
Lorentzian spacetime. The integration is well-defined so long as
$2 \hat{\ep} s_c^{\rm UV} > \ep >0$ is retained.


The second line in \eqn{zl} combines contributions of the
SL(2)$_k$/U(1), ${\cal M}$, and the worldsheet ghosts. The ghost
contribution $\eta(q_c)^2 \frac{\eta(q_c)}{\th_3(q_c)}$ is seen to
cancel out the contribution of longitudinal oscillators. Thus, the
amplitude simplifies to
\begin{eqnarray}
&& \hskip0.5cm Z_{\rm cylinder} \nn &=& i\frac{\alpha'}{2} \int_{s_c^{\rm
UV}}^{s_c^{\rm IR}} \rmd s_c \int_{0}^{\infty} \rmd p
\int_{-\infty}^{\infty} \rmd \om_L \, \frac {\sinh
\left(\frac{2\pi}{Q}\sqrt{\frac{\alpha'}{2}}p \right)
q_c^{\frac{1}{4}\alpha'(p^{2}-(1-i\hat{\ep})^2\om_L^{2})} \cdot
Z_{\cM}^{(c)}(q_c) } {\left[\cosh\left(\frac{2\pi}{Q} \sqrt{\frac{\alpha'}{2}}
\om_L\right) +\cosh\left(\frac{2\pi}{Q}\sqrt{\frac{\alpha'}{2}} p
\right)\right] \sinh(\pi Q \sqrt{\frac{\alpha'}{2}} p)}~. \nn
%
&&
\label{L cylinder closed}
\end{eqnarray}

We now modular transform \eqn{L cylinder closed} to the open string
channel.
Define again the open string modulus as $q_o=e^{-2\pi i \tau_o}$,
where $\tau_o = s_o -  i \ep$ and $s_o = 1/s_c$. Using the Fourier
transform identity:
\begin{eqnarray}
 && \int_{-\infty}^{\infty}\rmd x\,
\frac{\sin(\pi a x)}{\sinh(\pi x)} e^{-2\pi i kx} = \frac{\sinh (\pi
a)}{\cosh(2\pi k)+ \cosh (\pi a)} ~, \qquad
(\left|\mbox{Im}\,a\right| < 1) \label{FT formula}
\end{eqnarray}
we then obtain
\begin{eqnarray}
&& \hskip+0.5cm Z_{\rm cylinder} = \nn
&& \hskip-0.8cm \frac{i\alpha'}{4} \int_{s_o^{\rm UV}}^{s_o^{\rm IR}}
\frac{\rmd s_o}{s_o} \int_{-\infty}^{\infty}\rmd p'
\int_{-\infty}^{\infty}\rmd \om_L'\, \frac {\sinh
\left(\frac{2\pi}{Q} \sqrt{\frac{\alpha'}{2}} \om'_L \right)
q_o^{\frac{1}{4}\al'(p^{'2}-(1+i\hat{\ep}')^2\om_L^{'2})} \cdot
Z_{\cM}^{(o)}(q_o) } {\left[\cosh\left(\frac{2\pi}{Q}\sqrt{\frac{\alpha'}{2}}
\om'_L \right) +\cosh\left(\frac{2\pi}{Q}\sqrt{\frac{\alpha'}{2}}p'
\right)\right]\sinh(\pi Q \sqrt{\frac{\alpha'}{2}} \om_L')}.\,\,\,\,\,\,
\nn
%
&& \label{L cylinder open}
\end{eqnarray}
Again
$s_o^{\rm UV} \equiv 1/s_c^{\rm IR}$, $s_o^{\rm IR}\equiv
1/s_c^{\rm UV}$ are the cut-off's
and the expression \eqn{L cylinder
open} is well-defined so long as
$2 \hat{\ep}' s_o^{\rm UV}
> \ep$.

\subsection{analytic continuation}
We shall now analytically continue both the spacetime and the
worldsheet to the Euclidean signature. We have to carefully make the
continuation so that keeping the original amplitude \eqn{L cylinder
open} unchanged (up to cut-off's). As in the previous section, we
should first Wick rotate in spacetime $\om'_L\, \rightarrow\,
e^{i(\frac{\pi}{2}-0)} \om'_L$ with $\om'_L = i \om'$ $(\om' \in
\br)$, and then rotate the worldsheet $s_o \, \rightarrow \, - i
t_o$ ($t>0$).
%
%
%
We shall omit the cutoff's from now on.
We reach the expression
\be Z_{\rm cylinder} = Z_{\rm naive} + Z_{\rm pole}~, \ee
where the first part is the contribution from naive continuation,
while the second parts originates from the poles passed over by the
rotated contour: $\om'_L\, \rightarrow\, e^{i(\frac{\pi}{2}-0)}
\om'_L$. See Figure 2.
\vskip0.5cm
\begin{figure}[htbp]
    \begin{center}
    \includegraphics[width=0.6\linewidth,keepaspectratio,clip]
      {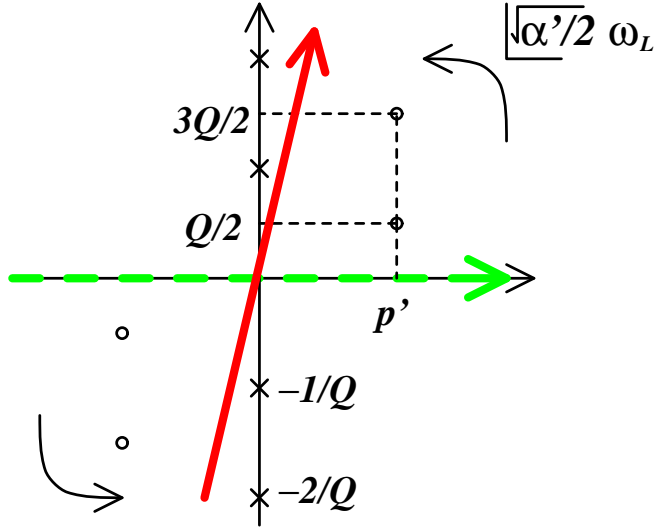}
    \end{center}
    \caption{\sl The $\om_L$-integral with Lorentzian contour (broken line)
    and the Euclidean contour (solid line).}
    \label{contour2}
\end{figure}
\vskip0.5cm
The first part $Z_{\rm naive}$ is given by
\be && \hspace{+0.2cm} Z_{\rm naive} = \nn
&& \hskip-1cm \int_{0}^{\infty} \frac{\rmd t_o}{t_o}
\int_{-\infty}^{\infty}\!\!\rmd p'
\int_{(1-i0)\mathbb{R}}\!\!\!\!\!\rmd \om'\, \frac {-
\frac{1}{4}\alpha' \sin \left(\frac{2\pi}{Q}
\sqrt{\frac{\alpha'}{2}}\om' \right)
q_o^{\frac{1}{4}\alpha'(p^{'2}+\om^{'2})} \cdot Z_{\cM}^{(o)}(q_o) }
{\left[\cos\left(\frac{2\pi}{Q}\sqrt{\frac{\alpha'}{2}} \om' \right)
+\cosh\left(\frac{2\pi}{Q} \sqrt{\frac{\alpha'}{2}} p'\right)\right]
\sin(\pi Q \sqrt{\frac{\alpha'}{2}} \om')}~.  \quad \label{Z naive}
%
\ee
The second part $Z_{\rm pole}$ arises from the poles located at
\begin{eqnarray}
 && \sqrt{\frac{\alpha'}{2}}\om'_L =
\left\{
\begin{array}{ll}
 \sqrt{\frac{\alpha'}{2}}|p'| + {iQ}
 \left(m+\frac{1}{2}\right) & ~~ m\in \mathbb{Z}_{\geq 0}  \\
 -\sqrt{\frac{\alpha'}{2}}|p'| + {iQ}
 \left(m+\frac{1}{2}\right) & ~~ m\in \mathbb{Z}_{<0 }
\end{array}
\right.
\end{eqnarray}
whose residues are (after taking the open string channel modulus
Euclidean, $q_o=e^{-2\pi t}$)
\begin{eqnarray}
 && \frac{i}{2} \sqrt{\frac{2}{\alpha'}}\cdot
\frac{Q}{2 \pi} \frac{e^{\pm i\pi t Q (2m+1)\sqrt{\frac{\alpha'}{2}} |p'| -
\pi t Q^2\left(m+\frac{1}{2}\right)^2}} {\sinh \pi Q \left(\pm
\sqrt{\frac{\alpha'}{2}} |p'|+iQ \left(m+\frac{1}{2}\right)\right)} ~.
\end{eqnarray}
We thus obtain
\begin{eqnarray}
 &&
Z_{\msc{pole}} = \int_0^{\infty} \frac{\rmd t_o}{t_o}\, \left\lb
\sum_{m\geq 0} \int_{0}^{\infty}\sqrt{\frac{\alpha'}{2}}\rmd p'
 -\sum_{m<0} \int_{-\infty}^{0}\sqrt{\frac{\alpha'}{2}}\rmd p'
\right\rb \,  
{e^{i\pi t Q (2m+1)\sqrt{\frac{\alpha'}{2}} p' - \pi t
Q^2\left(m+\frac{1}{2}\right)^2}}
\nn && \hskip3cm \times 2\pi i \cdot \frac{iQ}{4 \pi} \left\lb
\frac{1}{\sinh \pi Q \left(\sqrt{\frac{\alpha'}{2}} p'+iQ
\left(m+\frac{1}{2}\right)\right)} + (p' \leftrightarrow -p')
\right\rb Z_{\cM}^{(o)}(q_o)\nn
&& \hspace{1cm} = - 2Q  \int_0^{\infty} \frac{\rmd t_o}{t_o}\,
\sum_{m=0}^{\infty} \int_{0}^{\infty}\sqrt{\frac{\alpha'}{2}}\rmd
p'\, \frac{e^{i\pi t Q (2m+1)\sqrt{\frac{\alpha'}{2}} p' - \pi t
Q^2\left(m+\frac{1}{2}\right)^2}} {\sinh \pi Q \left(
\sqrt{\frac{\alpha'}{2}} p'+iQ \left(m+\frac{1}{2}\right)\right)}
\cdot Z_{\cM}^{(o)}(q_o) ~. \nn && \label{Z pole}
\end{eqnarray}

We thus obtained manifestly convergent open string channel
expressions \eqn{Z naive}, \eqn{Z pole} for the cylinder amplitude
in Lorentzian signature of the spacetime.

\subsection{optical theorem at work}
With the Lorentzian (in spacetime) cylinder amplitude \eqn{Z naive},
\eqn{Z pole} available, we now apply the unitarity and obtain total
emission number $\overline{\cal N}$ via imaginary part of $Z_{\rm
cylinder}$. In the analysis of \cite{OR} only the naive contribution
$Z_{\msc{naive}}$ was considered. Taking the imaginary part picks up
infinite poles located at the real $\om'$-axis (the imaginary
$\om'_L$-axis), depicted in Figure 2. Their contributions yield
\begin{eqnarray}
  && \mbox{Im}\, Z_{\msc{naive}} \nn
  & =& - \frac{1}{2}
\int_{0}^{\infty} \frac{\rmd t_o}{t_o} \int_{-\infty}^{\infty} \sqrt{\frac{\alpha'}{2}}\rmd
p' \sum_{\stackrel{n\neq 0}{n\in \bsz}} \, \pi \mbox{sgn}\,(n) \,
\frac{(-1)^n}{\pi Q} \frac {\sin\left(\frac{2\pi n}{Q^2}\right)
e^{-\pi t \left( \frac{\alpha'}{2} p^{'2}+\frac{n^2}{Q^2}\right)}}
{\cos\left(\frac{2\pi n}{Q^2}\right) +\cosh \left(\frac{2\pi
}{Q}\sqrt{\frac{\alpha'}{2}} p' \right)} \cdot Z_{\cM}^{(o)}(q_o) \nn &=&
-\sum_{n=1}^{\infty}\int_{0}^{\infty} \frac{\rmd t_o}{t_o}
\int_{-\infty}^{\infty} \sqrt{\frac{\alpha'}{2}} \rmd p' \, \frac{(-1)^n}{Q} \frac
{\sin\left(\frac{2\pi n}{Q^2}\right) e^{-\pi t \left(\frac{\alpha'}{2}
p^{'2}+\frac{n^2}{Q^2}\right)}} {\cos\left(\frac{2\pi n}{Q^2}\right)
+\cosh \left(\frac{2\pi}{Q}\sqrt{\frac{\alpha'}{2}}p'\right)} \cdot
Z_{\cM}^{(o)}(q_o)~, \label{ImZ 1st}
\end{eqnarray}
reproducing the first term in \eqn{ImZ final}.

We next evaluate the contribution from the pole contribution
$Z_{\msc{pole}}$ \eqn{Z pole}. As is easily seen, taking the
imaginary part just amounts to extending the integration region of
$p'$ in \eqn{Z pole}
to the whole real axis $(-\infty, \infty)$. By closing the
$p'$-contour in the upper half plane, we thus obtain
\begin{eqnarray}
&& \mbox{Im}\, Z_{\msc{pole}} \nn
&=&
 iQ \int_0^{\infty} \frac{\rmd t_o}{t_o}\,
\sum_{m=0}^{\infty} \int_{-\infty}^{\infty}\sqrt{\frac{\alpha'}{2}}\rmd p'\, \frac{e^{i\pi t
Q (2m+1)\sqrt{\frac{\alpha'}{2}} p' - \pi t Q^2\left(m+\frac{1}{2}\right)^2}}
{\sinh \pi Q \left(\sqrt{\frac{\alpha'}{2}} p'+iQ
\left(m+\frac{1}{2}\right)\right)} \cdot Z_{\cM}^{(o)}(q_o) \nn
&=&
 2\pi i \cdot iQ
\int_0^{\infty} \frac{\rmd t_o}{t_o}\,   \sum_{m=0}^{\infty}
\sum_{\stackrel{n> Q^2\left(m+\frac{1}{2}\right)} {n\in \bsz_{>0}}}
\, \frac{(-1)^n}{\pi Q} e^{-\pi t n (2m+1)+\pi t
Q^2\left(m+\frac{1}{2}\right)^2} \cdot Z_{\cM}^{(o)}(q_o)
 \nn
&=& -2  \int_0^{\infty} \frac{\rmd t_o}{t_o} \,
\sum_{n=1}^{\infty} \sum_{m=0}^{\left\lb
\frac{n}{Q^2}-\frac{1}{2}\right\rb} \, (-1)^n e^{-\pi t n (2m+1)+\pi
t Q^2\left(m+\frac{1}{2}\right)^2} \cdot Z_{\cM}^{(o)}(q_o)~.
\label{ImZ pole}
\end{eqnarray}
In the last line, we exchanged order of the double summations. The
final result agrees perfectly with the total emission number
$\overline{\cal N}$ in \eqn{ImZ final} evaluated via direct
computation of the transition amplitudes in Euclidean worldsheet.

\section{Interpretations and Discussions}
In the previous sections, we studied spectral observables in causal
processes involving decay of unstable D-brane and rolling of
accelerated D-brane. The main result of this work is that
transformation of the total emission number $\overline{\cal N}$ and
the cylinder amplitude $Z_{\rm cylinder}$ from the closed string
channel to the open string channel require careful analytic
continuation on the worldsheet and that, unlike other results
claimed in the literatures, the analytic continuation we adopt gives
results consistent with the unitarity via the optical theorem
$\overline{\cal N} = \mbox{Im}\, Z_{\rm cylinder}$.
In particular, we
found that the cylinder amplitude consists in general of two parts
$Z_{\rm cylinder} = Z_{\rm naive} + Z_{\rm pole}$, and the second
part is crucial for ensuring the unitarity through its imaginary
part. While we dealt with decaying or rolling process of the
D-brane, the rules we developed ought to extend to other real-time
processes such as open string and D-brane dynamics in electric field
or plane-wave field background.

In this section, we highlights several impotant steps we noted in
establishing consistency between the channel duality and the
unitarity.

\subsection{imaginary D-instantons: decaying versus rolling}

Throughout this work, the strategy for recasting the closed string
emission spectra in open string channel was to expand the transition
probability ${\cal P}(\omega, {\bf p})$ in power series
of `imaginary D-instantons' \cite{MSY,LLM,GIR}, viz.
contributions of localized states at time $2 \pi
i \alpha' W(m,n)$ for decaying D-branes and at time $(2 \pi i / Q) n
\sqrt{\frac{\alpha'}{2}}$ for rolling D-branes, respectively.

A crucial difference we noted for the rolling D-brane in NS5-brane
background, $Q < \sqrt{2}$, that weight of the $n$-th imaginary
D-instanton, $a_n(p)$, is a non-trivial function of $p$. We
emphasized above that the momentum dependence came about because
accelerated D-brane rolls in the two-dimensional subspace
$\mathbb{R}_t \times \mathbb{R}_\phi$. Being process dependent, it
could be that, in general, {\em the weights are exponentially
growing functions of momentum, and their Fourier transformations are
not necessarily well-defined.}
This was indeed the case for the rolling D-brane case. We thus
prescribed the Fourier transform of the D-instanton weight by
analytic continuation via a deformed integration contour. The
prescription then yielded in the open string channel the
contribution $Z_{\rm pole}$ beyond the naive one $Z_{\rm naive}$.
Moreover, whereas the naive contribution is always ultraviolet
finite, the pole contribution exhibited ultraviolet divergence.
Since $\overline{\cal N}$ (or higher spectral moment) is ultraviolet
divergent, we concluded that the presence of ultraviolet divergent
$Z_{\rm pole}$ is crucial for consistency with the unitarity and the
channel duality.

From mathematical viewpoint, we found that the pole contribution
$Z_{\rm pole}$ in \eqn{ImZ final} is present in so far as we adopt
mathematically well-posed prescription of the Fourier transform.
>From physics viewpoint, we can also argue that the first term
$Z_{\rm naive}$ by itself cannot be the correct answer and the
second term $Z_{\rm pole}$ ought to dominate over the first one.

In the range $Q < \sqrt{2}$, it is easy to see that
$\mbox{Im}\,Z_{\rm naive}$ can take a negative value if we tune the
value $Q$ suitably within this range. If $Z_{\rm naive}$ is all
there is for the cylinder amplitude, the negative value of its
imaginary part contradicts with the fact that the total emission
number $\overline{\cal N}$ is positive by definition. Moreover, for
$Q=\sqrt{2/k}$ $k\in \mathbb{Z}_{>1}$, which corresponds to rolling
D-branes in $k$ coincident NS5 backgrounds, we observe that the
first term $Z_{\rm naive}$ vanishes identically since the integrand
vanishes. The above observations indicate that extra contribution
ought to be present to the cylinder amplitude beyond the naive
contribution, $Z_{\rm naive}$.

On other other hand, we do not have any contradiction of the
cylinder amplitude with the unitarity once the contribution $Z_{\rm
pole}$ is taken into account. This is because $Z_{\rm pole}$ is
dominant (generically divergent) over $Z_{\rm naive}$ and always
positive. We conclude that our prescription for the cylinder
amplitude renders the total emission number, as extracted from the
optical theorem as $\overline{\cal N} = \mbox{Im}\,Z$ always positive
and well-defined.

The situation is in sharp contrast to that for decaying D-brane
case. There, as recapitulated in section 2, the D-instanton weights
were constant ($a_{n,m} = 1$), so the issue of Fourier transform was
void from the outset. Again, as explained in section 2, the momentum
independence came about because unstable D-brane decays at rest (or
trivially Lorentz boosted). The situation in NS5-brane phase $Q <
\sqrt{2}$ is also in contrast to that in the 'fundamental string
phase' \cite{GKRS}, $\sqrt{2} < Q \leq 2$, or in `out-going'
radiation in nonextremal NS5-brane background (which involves
two-dimensional black hole geometry) \cite{NRS}. For these, the
leading weight $a_1(p)$ is a bounded function and have well-defined
Fourier transformation. So, there does not arise any extra
contribution beyond $Z_{\rm naive}$. We thus obtain via optical
theorem an ultraviolet finite total emission number~\footnote
  {Even in the deep stringy phase $\sqrt{2}< Q \leq 2$, $a_n(p)$ is
  exponentially divergent for sufficiently large $n$.
  So, the formula given in \cite{OR} have to be still
  corrected. However, only the $n=1$ term
  could cause the Hagedorn divergence as noted above.
   Hence, this correction does
   not modify ultraviolet behavior of the emission number density.}.

In the previous work \cite{NRS}, we also noted that the first
D-instanton weight $a_1(p)$ is identifiable with the `grey body
factor' $\sigma(p)$ in the total emission number $\overline{\cal
N}$. There, the identification was based on saddle-point analysis
valid at large mass $M \rightarrow \infty$ in the closed string
channel. The present result in the open string channel, where the
leading ultraviolet divergence arises from the weight $a_1(p)$, then
supports the identification~\footnote{Footnote 3 of \cite{OR} claims
the saddle-point approximation used in our earlier works \cite{NST,
NPRT, NRS} is invalid. We disagree with their claim: the relevant
integral is of the type
\be \int^\infty \rmd p \, \exp \Big[-M f\Big({p \over
M}\Big)\Big]. \nonumber \ee
As $M \rightarrow \infty$, the saddle point approximation is well
justified in so far as
$$
f(p_*/M) \sim \cO(1)~, ~~~ f^{(2)}(p_*/M) >0~, ~~~ f^{(2n)}(p_*/M)
\sim \cO(1)~  ~~ (n \geq 1)~, ~~~ (p_*~:~ \mbox{saddle})~,
$$
and this was indeed so in our previous works \cite{NST, NPRT,
NRS}.}.

\subsection{comparisons}

From our analysis, it became clear the reason why \cite{KLMS}
obtained the correct result for the decay of unstable D-brane is
because the contour rotation in Fourier transform did not encounter
any pole (since the D-instanton weights $a_n(p)$ were
$p$-independent constants), and the naive manipulation yielded the
correct result. In \cite{OR}, the prescription of \cite{KLMS} was
taken literally also for the rolling of accelerated D-brane. It was
then concluded that $Z_{\rm naive}$ refers to the total cylinder
amplitude. We showed throughout this work that this is incorrect
since it overlooked the pole contribution $Z_{\rm pole}$. After all,
only after taking this extra contribution into account, we showed
that the cylinder amplitude is consistent with the channel duality
and the unitarity.

Finally, we find it illuminating to understand why $\overline{\cal
N}$ exhibited Hagedorn divergence in the two-dimensional string
theory studied in \cite{KMS}, whereas it is ultraviolet finite in
the linear dilaton background studied in \cite{KLMS} in
two-dimensional spacetime (that is, $c_{\msc{eff}}=0$). The reason
is because the boundary wave function (D-brane transition amplitude)
of the former has non-trivial $p$-dependence that exponentially
diverges, whereas the latter does not.

\subsection*{Acknowledgement} We thank David J. Gross and Jung-Tay
Yee for useful discussions, and acknowledge Kazumi Okuyama
and Moshe Rozali for correspondences.
YN's work was supported in part by JSPS
Research Fellowship for Young Scientists. SJR's work was supported
in part by the KOSEF SRC Program "Center for Quantum Spacetime"
(R11-2005-021).
YS's work was supported in part by the Ministry of Education,
Culture, Sports, Science and Technology of Japan.

\end{document}